\documentclass{emulateapj}

\newcommand{\be}{\begin{equation}}
\newcommand{\ee}{\end{equation}}
\newcommand{\nh}{N_{\rm H}}
\newcommand{\pdot}{\dot{P}}
\newcommand{\edot}{\dot{E}}
\newcommand{\chan}{{\sl Chandra}}
\newcommand{\xmm}{{\sl XMM-Newton}}

\begin{document}

\title{ Detection of X-ray emission from the very old pulsar J0108--1431
  }

\author{G.\ G.\ Pavlov$^1$, O.\ Kargaltsev$^{1,2}$, J.\ A.\ Wong$^{1}$, and G.\ P.\ Garmire$^1$ }
\affil{$^1$Dept.\ of Astronomy and Astrophysics, Pennsylvania State University, PA 16802;
pavlov@astro.psu.edu\\
$^2$Dept.\ of Astronomy, University of Florida, Bryant Space Science
Center, Gainesville, FL 32611; oyk100@astro.ufl.edu}
%\author{G.\ G.\ Pavlov\altaffilmark{1}
%\altaffiltext{1}{
%Dept.\ of Astronomy and Astrophysics, Pennsylvania State University, PA 16802; 
%pavlov@astro.psu.edu}, 
%O.\ Kargaltsev$^{1,}$
%\altaffilmark{2}\altaffiltext{2}{
%Dept.\ of Astronomy, University of Florida, Bryant Space Science
%Center, Gainesville, FL 32611; oyk100@astro.ufl.edu}, 
%J.\ A.\ Wong$^{1}$,
%\altaffilmark{1}, 
%and G.\ P.\ Garmire$^1$ }
%\altaffilmark{1}}

\begin{abstract}
PSR J0108--1431 is a nearby,
 170 Myr old, very faint radio pulsar near the ``pulsar
death line'' in the $P$-$\pdot$ diagram. 
We observed the pulsar field with the {\sl Chandra X-ray Observatory} and 
detected a point source 
(53 counts in a 30 ks exposure; energy flux 
$(9\pm 2)\times 10^{-15}$ ergs cm$^{-2}$ s$^{-1}$
in the 0.3--8 keV band) close to the radio pulsar position.
Based on the large X-ray/optical flux ratio at the
X-ray source position, we conclude
that the source is the X-ray counterpart of PSR J0108--1431.
The pulsar spectrum can be described by a power-law model
with photon index $\Gamma\approx2.2$
and luminosity
$L_{\rm 0.3-8\, keV}\approx 2\times 10^{28} d_{130}^2$ ergs s$^{-1}$, 
or by a blackbody model with
the temperature $kT\approx0.28$
keV and 
bolometric luminosity $L_{\rm bol} \approx 1.3\times
10^{28} d_{130}^2$
ergs s$^{-1}$, for a plausible hydrogen column density $\nh = 7.3\times
10^{19}$ cm$^{-2}$ ($d_{130}=d/130\,{\rm pc}$). 
The pulsar converts $\sim 0.4\%$ of
its spin-down power into
the X-ray luminosity, i.e., its X-ray efficiency is higher than 
for most younger pulsars.
From the comparison of the X-ray position with the previously
measured radio positions, 
we estimated the pulsar proper motion 
of 0.2 arcsec yr$^{-1}$ ($V_\perp\approx
130 d_{130}$ km s$^{-1}$), in the south-southeast direction.

\end{abstract}


\section{Introduction}

Because of the observational selection, most of radio pulsars detected 
in X-rays are much younger and more powerful than ``typical'' ones, 
and they cannot be considered
as a representative sample of the whole radio pulsar population.
Therefore, to understand the evolution of X-ray pulsar properties,
 it is important to expand the current sample toward lower spin-down
powers, $\edot$, and larger spin-down ages, $\tau$.
X-ray emission from such low-powered, old pulsars is expected
to be composed of
nonthermal (magnetospheric) emission and 
thermal emission  from the heated
polar caps around the neutron star (NS) magnetic poles. (Old NSs are too cold
to observe thermal emission from the bulk of the NS surface in X-rays;
see, e.g., Yakovlev \& Pethick 2004.)
The magnetospheric X-ray emission is routinely seen in both
young and old pulsars.
It is characterized by a power-law (PL) spectrum, with a typical slope
(photon index) $1\lesssim \Gamma \lesssim 2$, and relatively sharp pulsations.
The X-ray emission from polar caps, heated by 
relativistic particles accelerated in the
pulsar magnetosphere, 
is expected to be most efficient in old pulsars
(e.g., Harding
\& Muslimov 2001, 2002).
It should show a thermal spectrum, with a temperature $kT\sim 0.1$--0.3 keV,
and smoother pulsations.
Since both the magnetospheric and thermal components are powered
by the NS spin-down, their luminosities
are expected to be proportional to the spin-down power,
$L_X = \eta_X \edot$, where $\eta_X< 1$ is the pulsar X-ray efficiency.

As of February 2008,
\chan\ and \xmm\ observations of eight old, low-powered pulsars
 ($1\,{\rm Myr}<\tau < 20\,{\rm Myr}$, $31.9 < \log\edot < 33.6$)
have been reported, of which seven were detected
(Becker et al.\ 2004, 2005, 2006;
Zavlin \& Pavlov 2004; Zhang et al.\ 2005; Tepedelenlio\u{g}lu
\& \"{O}gelman 2005; Kargaltsev et al.\ 2006; Hui \& Becker 2007;
Misanovic et al.\ 2008)\footnote{After submitting this paper, Gil
et al.\ (2008) reported \xmm\ observations of two other old pulsars,
of which one, PSR B0834+06,
 was detected and showed properties similar to the previously
detected old pulsars.}.
Most of them showed
rather faint X-ray emission, such that it was hard to accurately characterize
their spectra and detect pulsations.
The spectra of all the detected pulsars can be
satisfactorily  fitted with a PL model,
but the PL slopes, $\Gamma\sim 2$--4, are, on average,
 steeper than those found in young
pulsars, and the hydrogen column densities, $N_{\rm H}$,
often exceed those
estimated from the pulsar dispersion measure at the usual assumption of
10\% ionization of
the interstellar medium (ISM). (An extreme example is PSR B1929+10,
for which the $\nh$ found from the PL fit corresponds to  0.4\% 
of ISM ionization; Misanovic et al.\ 2008).
Some of the spectra can be fitted with a 
blackbody (BB) model, but the emitting
areas are usually much smaller than the conventional areas of polar caps.
The spectra can be also described by a two-component, PL+BB, model,
but usually with poorly constrained parameters.
The measured X-ray
efficiencies
of old pulsars show a large
scatter, $\eta \sim 10^{-4}$--$10^{-2}$ in the 1--10 keV band,
 but, on average, they seem to be higher than those for young
pulsars (Kargaltsev et al.\ 2006).
Pulsations were firmly detected only for
three brighter pulsars, B1929+10, B0628--28, and B0950+08, 
with a hint of energy
dependence in the latter case,
but their
analysis was not very conclusive (e.g., Becker et al.\ 2004 and
Zavlin \& Pavlov 2004 came to quite different conclusions about the
possible polar cap contribution to the emission from B0950+08). At this point,
our working hypothesis is that the emission from old pulsars possibly
includes both the magnetospheric
and polar cap components, such that the magnetospheric component dominates
at higher energies, $\gtrsim 1$--2 keV, and has about the same
PL slope and efficiency as in young pulsars. However, we cannot rule out that
old pulsars are purely magnetospheric emitters, with spectra softer,
and efficiencies higher, than those of young pulsars.
To check which of these two hypotheses is correct,
not only deeper observations of X-ray brightest old pulsars are warranted,
but also  the number of observed pulsars should be increased.
Particularly important would be X-ray observations of pulsars
even older and less powerful than those
in the currently observed sample.

\begin{figure}[t]
% \centering
\hspace{-0.5cm}
\includegraphics[width=2.7in,angle=90]{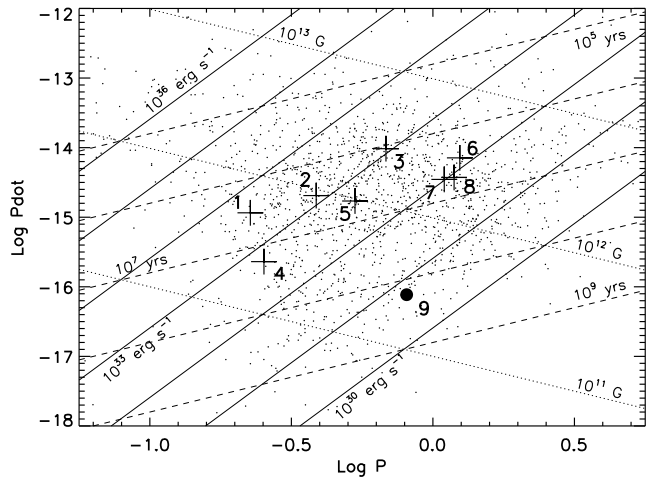}
\caption{Portion of the
$P$-$\dot{P}$ diagram for pulsars from the ATNF Pulsar Catalogue
(Manchester et al.\ 2005). The eight old, low-powered pulsars
previously observed with {\sl Chandra} and {\sl XMM-Newton} are marked
by crosses;  J0108 is marked by the filled circle.
The numbers 1 through 9 near the marked pulsars correspond to those
in Fig.\ 9. Notice that
J0108 is located close to the bottom-right boundary of the ``swarm''
of the pulsar population, generally known as the ``pulsar death line''.
}
\label{p-pdot}
\end{figure}

A very interesting target for such observations is PSR J0108--1431
(hereafter J0108),
a very old, non-recycled pulsar (period $P=0.808$ s,
spin-down age $\tau = P/2\dot{P}=
166$ Myr, spin-down power $\dot{E}=5.8\times 10^{30}$ ergs s$^{-1}$,
surface magnetic field $B=2.5\times 10^{11}$ G),
discovered by Tauris et al.\ (1994).
Based on its
dispersion measure, DM = 2.38 pc cm$^{-3}$
(D'Amico et al.\ 1998),
the smallest known for any radio
pulsar,
and the Galactic coordinates, 
$l=140.9^\circ$ and $b=-76.8^\circ$,
the distance to the pulsar is $d=130$ pc,
according to the
model for Galactic electron density distribution by Taylor \& Cordes (1993)
($d=180$ pc in the model by Cordes \& Lazio 2002).
J0108 is not only among the nearest NSs to Earth,
but it also has the second lowest known radio luminosity, 
0.15 mJy kpc$^2$ at
400 MHz (for $d=130$ pc), 
and its position in the $P$-$\dot{P}$ diagram is very close
to the ``pulsar death line'' (see Fig.\ \ref{p-pdot}).
Mignani et al.\ (2003) carried out 
deep optical
observations of the J0108 field with the Very Large Telescope (VLT)
and reported upper limits
on its fluxes in several bands (e.g., $V>28.0$, $B>28.6$).

Obviously, 
even a mere detection of such an
extremely old, ``almost dead'' pulsar in X-rays
would allow one to probe the pulsar properties in the regime where no
other pulsar has been detected outside the radio range. Therefore,
we observed J0108 with the \chan\ X-ray Observatory.
We describe the observation and data analysis in \S\,2, and discuss
implications of our results in \S\,3.
%
%%%%%%%%%%%%%%%%%%%%%%%%%%%%%%%
\section{Observations and data analysis}
%%%%%%%%%%%%%%%%%%%%%%%%%%%%%
The field of J0108 was observed by {\sl Chandra} on 2007 February 5
(MJD 54,136) using 
the Advanced CCD Imaging Spectrometer (ACIS). 
The observation was carried out in Very Faint mode, and the 
target was imaged on the ACIS-S3 chip, 
$7''$ off-axis.
The detector was operated in Full Frame mode, which provides
time resolution of 3.24 s.
The total scientific exposure time was 30,090 s. As
no bad-time intervals (e.g., particle background flares) were found
in the data, the whole exposure was used for the analysis.
Data reduction was done with
the Chandra Interactive Analysis Observations (CIAO) software
(ver.\ 3.4).
For the analysis, we used the data reprocessed on 2007 August 5
(ASCD ver.\ 7.6.11, CALDB ver.\ 3.4.1)\footnote{This reprocessing
corrected for a systematic
aspect offset of $\approx0.4''$, believed to be the result of a changed 
thermal environment in early 2007.}.
We removed the pipeline pixel randomization
and used the energy range of 0.3--8 keV to minimize the particle
background.
We used 
XSPEC ver.\ 11.3.2ag for spectral fits.

\begin{figure}[h]
 \centering
\includegraphics[width=3.2in,angle=0]{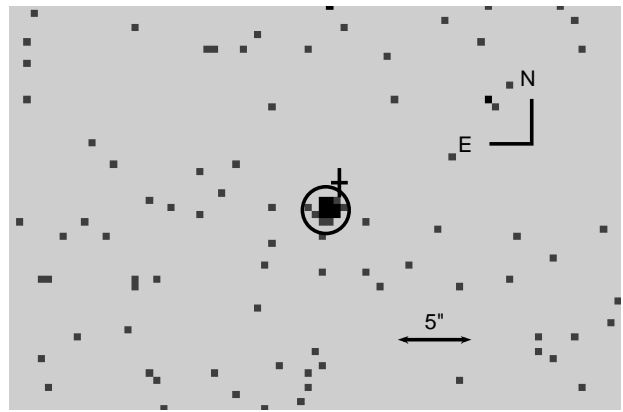}
\caption{Original ACIS image
in the 0.3--8 keV band (pixels size $0.492''$). The $1.6''$ radius circle
shows the source aperture from which the counts for spectral analysis
were extracted.
The cross shows the radio pulsar position from the ATNF Pulsar Catalogue
(Manchester et al.\ 2005).}
\label{ACIS_image}
\end{figure}

\subsection{Image}
Inspection of the ACIS image shows only one source in the vicinity of 
the radio pulsar position 
(see Fig.\ \ref{ACIS_image}).
 To measure the coordinates of this pointlike source, we used
the CIAO tasks {\em celldetect}, {\em wavdetect} and 
{\em vtdetect}, which yielded the same 
source position within
the centroiding uncertainty ($\approx0.05''$ for each of the coordinates):
\begin{equation}
\alpha  =  01^{\rm h}08^{\rm m}08.354^{\rm s}, \quad
\delta  = -14^\circ 31'50.38''.
\end{equation}
The astrometric uncertainty of this position can be estimated
from the empirical distribution of the radial offsets of X-ray positions
(aspect solutions) from the accurately known celestial locations for
a sample of point 
sources\footnote{See \S\,5.4 and
 Fig.\ 5.4 in the Chandra Proposers' Observatory 
Guide, ver.\ 10, at http://asc.harvard.edu/proposer/POG.}. 
In particular,
68\% of the 184 such sources imaged on the S3 chip within $2'$ from the
optical axis have offsets smaller than $0.21''=R_{0.68}$, 
and 90\% of the sources have offsets
$<0.44''=R_{0.90}$. For a two-dimensional Gaussian distribution with
equal standard deviations $\sigma$ for the (uncorrelated) right ascension
and declination offsets, we can estimate the standard deviation as 
$\sigma=R_a\left\{2\ln\left[(1-a)^{-1}\right]\right\}^{-1/2}$,
where $a$ is the probability to have a radial offset smaller than $R_a$.
This gives $\sigma=R_{0.68}/1.51=0.14''$
and $\sigma=R_{0.90}/2.15 = 0.205''$. The different values of $\sigma$
are likely due to the scarce statistics, and we can adopt the larger
one as a conservative estimate.
Adding the centroiding uncertainty in quadrature, we obtain the total
uncertainty $\sigma_\alpha=\sigma_\delta=0.21''$ for each of the coordinates.

The absolute astrometric position of the target could be improved, 
in principle, 
by cross-correlation of the ACIS positions of field X-ray sources with
the positions of their optical counterparts in catalogs with sufficiently
accurate astrometry, such as the 2MASS catalog.
However, we found only two X-ray sources with possible 2MASS counterparts.
One of these sources, located $3.3'$ northwest of the target, has too few
counts to measure its centroid with sufficient precision (centroiding
errors are $0.22''$ in right ascension and $0.11''$ in declination, while
the corresponding {\sl Chandra} $-$ 2MASS offsets are $0.22''$ and $-0.15''$).
As the other source, $3.7'$ southeast of the target, looks slightly extended 
(perhaps a double source) 
in X-rays, its centroid cannot be reliably measured (e.g., the positions
measured with {\em celldetect}
and {\em wavdetect} differ by $\approx 0.5''$). Therefore, the field sources
cannot be used for a reliable boresight correction.

The measured X-ray position differs by $2.10''$ from the radio pulsar position
in the ATNF Pulsar 
Catalogue\footnote{http://www.atnf.csiro.au/research/pulsar/psrcat} 
(Manchester et al.\  2005) derived from timing observations of
the pulsar (Hobbs et al.\ 2004),
and it differs by $1.16''$ from the radio-interferometric position
reported by Mignani et al.\ (2003).
These differences significantly exceed the absolute position
uncertainty of {\sl Chandra} source location and the reported uncertainties
of the radio positions (see Fig.\ 3).
Therefore, the source could be either a field object (a star or an AGN) or
the pulsar with an appreciable proper motion. 
Based on the lack of optical sources in the most conservative
{\sl Chandra} error circle, we will argue in \S\, 2.2 that only
the latter option is viable, and we will estimate the pulsar proper
motion in \S\, 3.1.

\begin{figure}[t]
 \centering
\includegraphics[width=3.4in,angle=0]{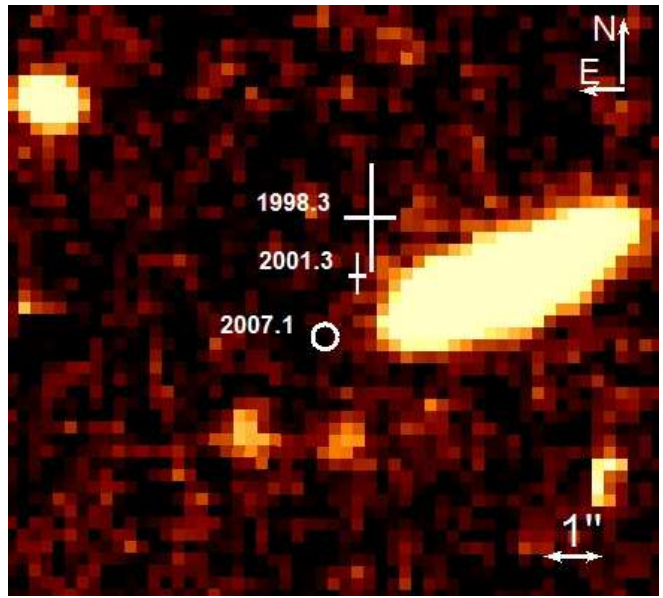}
\caption{$11''\times 10''$ VLT image of the J0108 field in the V band
(1200 s exposure taken with the FORS1 instrument of the VLT-Antu
in 2000 July 6-7; the reduced
VLT image was kindly provided by R.\ Mignani).
The position of the
detected X-ray source is shown by the $0.21''$ radius {\sl Chandra} error
circle (68\% confidence).
The crosses $1.16''$ and $2.10''$ north-northwest of the X-ray source
position correspond
to the radio pulsar positions from Mignani
et al.\ (2003) and Hobbs et al.\ (2004), respectively;
the sizes of the cross arms are the uncertainties of radio positions.
}
\end{figure}

\subsection{Spectral Analysis}
\begin{figure}[h]
% \centering
\hspace{-1cm}
\includegraphics[width=2.9in,angle=90]{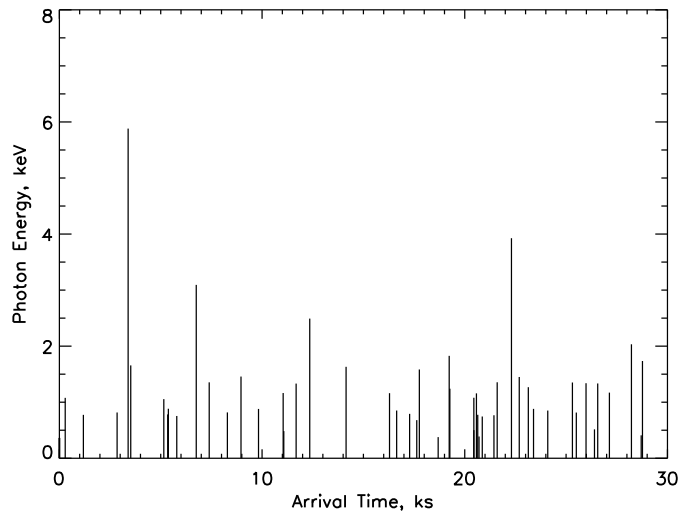}
\caption{Energies and arrival times for the 53 photons
extracted from the $1.6''$
aperture around the pulsar.
}
\label{energy-time}
\end{figure}

Using the CIAO's {\em psextract} task,
we extracted 53 counts 
from the circular aperture of $1.6''$ radius 
($\approx95\%$ encircled energy radius) around the centroid position
(Fig.\ \ref{ACIS_image}).
The background was taken from an annulus of $9'' < r < 18''$ 
around the pulsar.
The background region has 57 counts, which corresponds to
 0.60 counts when scaled to the source aperture. 
Thus, the source count rate is 
$1.74\pm0.24$ counts ks$^{-1}$, in the $1.6''$ aperture and 0.3--8 keV
energy range.
In Figure \ref{energy-time} we show the energies of all the 53 events
detected, which are within the range $0.3\,{\rm keV}<E<6\,{\rm keV}$,
 and the distribution of their arrival times
over the duration of the observation. The distribution of the arrival times
does not show statistically significant deviation from the Poisson
statistics,
i.e. the source does not show flares, which could be seen in X-ray
emission from a star or an AGN.

The observed X-ray flux and its variance can be estimated as
$F=T_{\rm exp}^{-1} \sum_{k=1}^N E_k/A_k$
and
$\sigma_F^2=T_{\rm exp}^{-2} \sum_{k=1}^N (E_k/A_k)^2$, 
where $T_{\rm exp}$ is the
exposure time, $N$ is the total number of detected events, $E_k$ is the energy
of $k$-th event, and $A_k=A(E_k)$ is the detector effective area at
energy $E_k$. 
For the source under consideration, 
the flux in the 0.3--8 keV band
is $F_{\rm 0.3-8\,keV}= (9.2\pm 1.9)\times 10^{-15}$ ergs cm$^{-2}$ s$^{-1}$.
This flux is much higher than the upper limits on the
optical fluxes within the $5\sigma$ {\sl Chandra}
error circle,
$F_V<2.6\times 10^{-17}$
and $F_B<2.4\times 10^{-17}$ ergs cm$^{-2}$ s$^{-1}$ in the V and B bands, 
respectively
(Mignani et al.\ 2003).
The large X-ray/optical flux ratio
[e.g., $R_V\equiv F_{\rm 0.3-3.5\, keV}/F_V > 230$, where
$F_{\rm 0.3-3.5\, keV}=(7.3\pm 1.2)\times 10^{-15}$ ergs cm$^{-2}$ s$^{-1}$]
means that the detected source cannot be an ordinary star
or an AGN, the usual field sources at high Galactic latitudes,
for which $R_V<0.3$ and $R_V<50$, respectively  
(see Fig.\ 1 in Maccacaro et al.\ 1988, and Table 1 and Fig.\ 1 in
Stocke et al.\ 1991).
The only known objects with $R_V>100$ are NSs (e.g., $R_V \sim 10^{2.5}$--$10^4$
for radio pulsars; see Zavlin \& Pavlov 2004).
 Therefore, 
we conclude
that the source should be the X-ray counterpart of PSR J0108--1431, whose
displacement from the previously measured pulsar positions can be
attributed to the proper motion of the pulsar.

Because of the small number of counts,
we used the $C$-statistic (Cash 1979) for spectral fitting,
without binning the count spectrum.
The results of spectral fits are presented in Table \ref{spectral}.
The errors of the 
fitting parameters in Table \ref{spectral} and below in the text
 are quoted at the 68$\%$ confidence level
for one interesting parameter, calculated with the XSPEC
{\em error} command. 

\begin{table}
\caption{Fits of the J0108 spectrum}
\vspace{2mm}
\begin{tabular}{lcccccc}
\tableline\tableline
Model & $N_{\rm H,20}$\tablenotemark{a} & $\Gamma$/$kT$\tablenotemark{b} & $\mathcal{N}_{-6}$/$A_\perp$\tablenotemark{c} & $F_{-14}^{\rm un}$\tablenotemark{d} & $C$\tablenotemark{e}
& $L_{28}$\tablenotemark{f} \\
\tableline
PL & $23^{+17}_{-12}$ & $3.4^{+1.0}_{-0.8}$ & $4.8^{+3.8}_{-1.8}$ & $2.9^{+0.7}_{-0.6}$ & 159.5
& 5.9 \\
PL & [0.73] & $2.20^{+0.24}_{-0.23}$ & $2.15^{+0.31}_{-0.28}$ & $1.1^{+0.3}_{-0.2}$ & 163.7
& 2.1 \\
PL & [2.0] & $2.28^{+0.24}_{-0.24}$ & $2.26^{+0.33}_{-0.30}$ & $1.1^{+0.3}_{-0.2}$ & 163.0 & 2.2 \\
BB & $0^{+5}_{-0}$ &
$281^{+37}_{-30}$ & $53^{+34}_{-21}$ & $0.6^{+0.2}_{-0.1}$ & 160.8
& 1.3 \\
BB & [0.73] &
$279^{+35}_{-28}$ & $53^{+32}_{-21}$ & $0.6^{+0.2}_{-0.1}$ & 161.0
& 1.3 \\
BB & [2.0] & $274^{+33}_{-27}$ & $62^{+36}_{-23}$ & $0.7^{+0.2}_{-0.1}$ & 161.3 & 1.4 \\
\tableline
\end{tabular}
\tablenotetext{a}{Hydrogen column density in units of $10^{20}$ cm$^{-2}$. Fixed values are in square brackets.}
\tablenotetext{b}{Photon index for the PL fits or temperature for the BB fits, in eV.}
\tablenotetext{c}{PL normalization in units of $10^{-6}$ photons cm$^{-2}$ s$^{-1}$ keV$^{-1}$ at 1 keV or apparent emitting area in m$^2$.}
\tablenotetext{d}{Unabsorbed flux in the 0.3--8 keV range, in units of $10^{-14}$ ergs cm$^{-2}$ s$^{-1}$.}
\tablenotetext{e}{Best-fit $C$ statistic value, for 526 bins used in the fit.}
\tablenotetext{f}{X-ray luminosity in the 0.3--8 keV band for the PL fits, or bolometric luminosity for the BB fits, in units of $10^{28}$ ergs s$^{-1}$.}
\label{spectral}
\end{table}

\begin{figure}[h]
% \centering
%\hspace{-0.5cm}
\includegraphics[width=2.5in,angle=90]{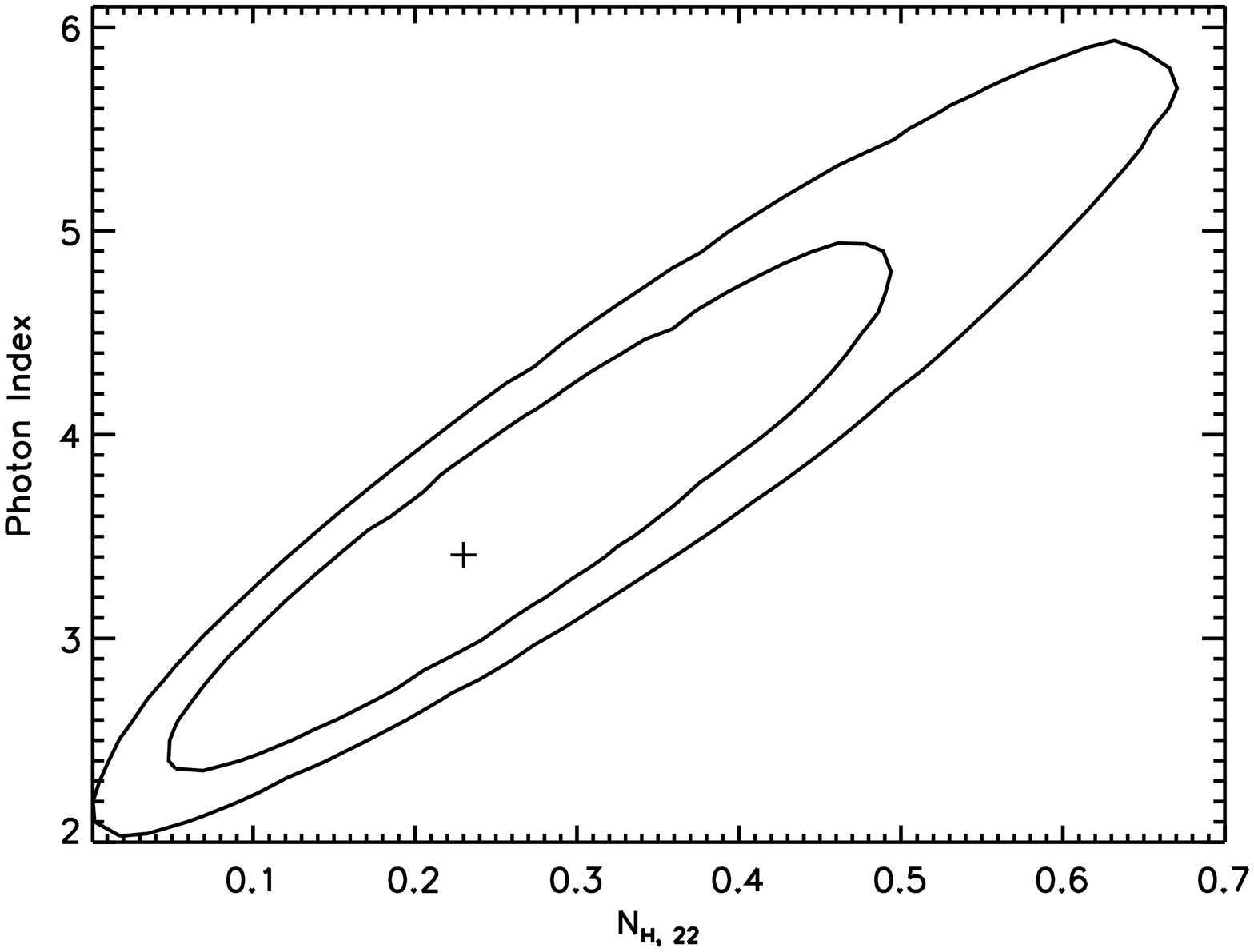}
%\hspace{-1.5cm}
\includegraphics[width=2.5in,angle=90]{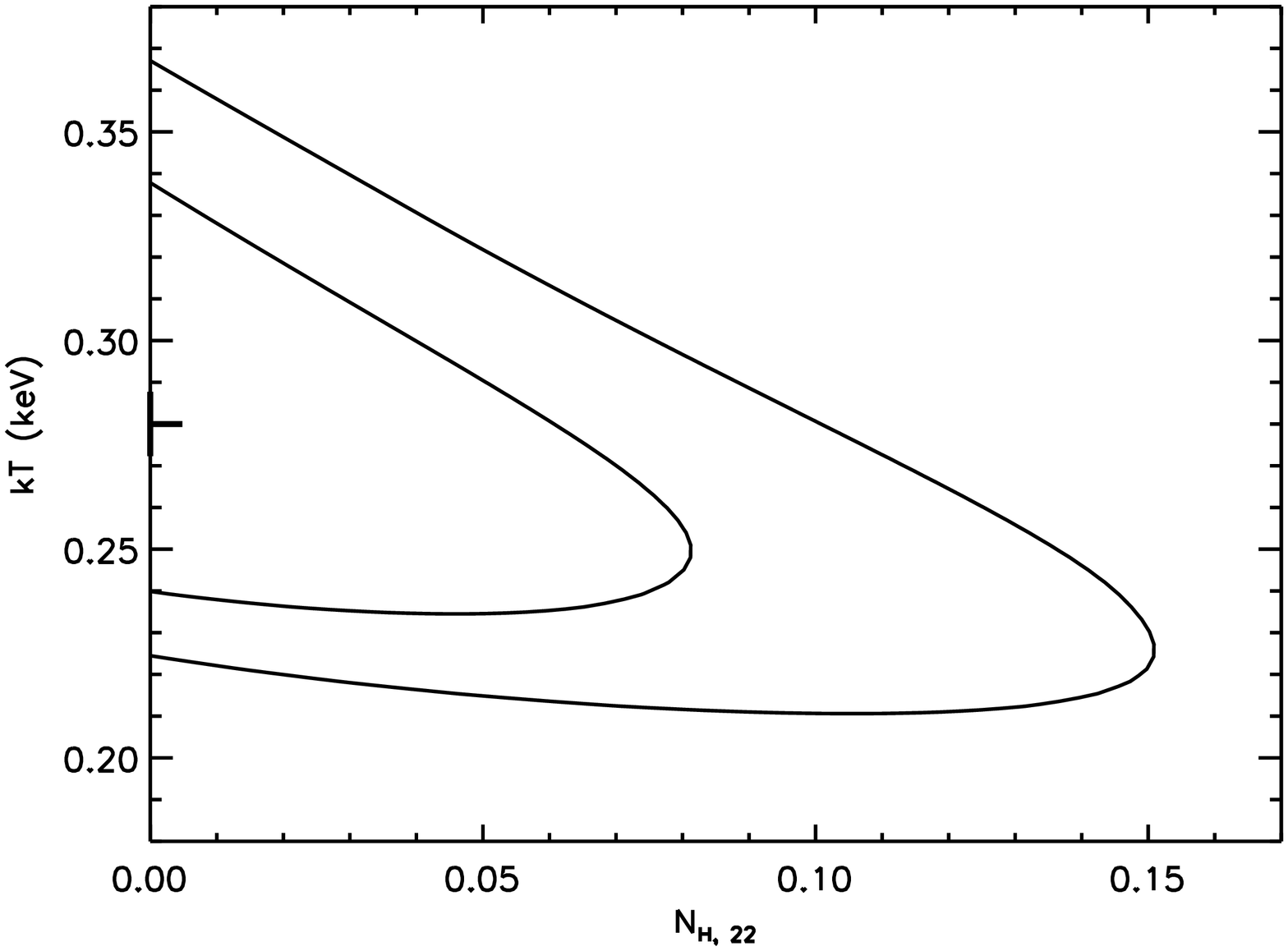}
\caption{68\% and 90\% confidence contours in the $\nh$-$\Gamma$ plane
for the PL model ({\em top}) and $\nh$-$kT$ plane for the BB model
({\em bottom}). The contours are obtained with the normalization
parameters fitted at each point of the grids.
}
\label{nh-gamma}
\end{figure}

First, we fit the spectrum with the absorbed PL model, 
with all the model parameters allowed to vary. 
Because of the small number of counts,
the fit is poorly constrained (see Fig.\ \ref{nh-gamma}, {\em top}).
Moreover, the best-fit hydrogen column density obtained in this fit,
$\nh= 2.3
\times10^{21}$ cm$^{-2}$, is unrealistically large,
at least a factor of 10
larger than the total Galactic neutral
hydrogen column density
in the direction of J0108: 
$N_{\rm HI} \approx
1.8\times10^{20}$ cm$^{-2}$ (Dickey and Lockman 1990) or
$2.1\times 10^{20}$ cm$^{-2}$ (Kalberla et al.\ 2005).
The best-fit photon index obtained  in this fit,  $\Gamma \approx 3.4$,
is considerably higher than those usually observed from radio pulsars.

\begin{figure}[h]
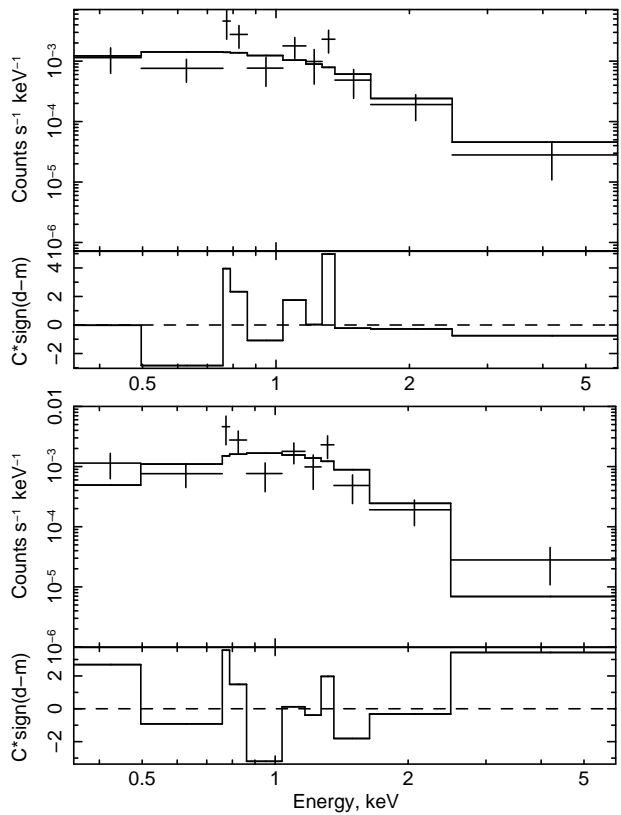

% \centering
\includegraphics[width=2.0in,angle=270]{f6a.ps}
\includegraphics[width=2.2in,angle=270]{f6b.ps}
\caption{
Binned count-rate spectra for
the data and the best-fit absorbed PL ({\em top}) and BB ({\em bottom})
 models
at fixed $\nh= 7.3\times10^{19}$ cm$^{-2}$ (see Table \ref{spectral}).
The counts in the data are binned to 5 counts per energy
bin (except for the last bin that contains 3 counts).
The lower parts of the panels show the contributions of
the spectral bins to
the $C$-statistics ($d-m$ is the difference between the data counts and
model counts in an energy bin).
}
\label{count_spectra}
\end{figure}

\begin{figure}[h]
% \centering
\hspace{-0.6cm}
\includegraphics[width=2.7in,angle=90]{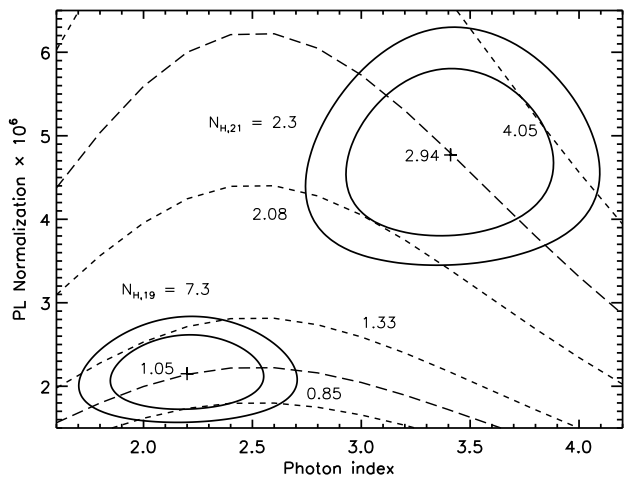}
\caption{68\% and 90\% confidence contours
in the $\Gamma$-$\mathcal{N}$ plane for the absorbed PL models with
$\nh$ values fixed at $2.3\times 10^{21}$ and
$7.3\times10^{19}$ cm$^{-2}$.
The PL normalization $\mathcal{N}$  is in units of $10^{-6}$
photons  keV$^{-1}$ cm$^{-2}$ s$^{-1}$.  The dashed lines
are the lines of constant unabsorbed
flux in the 0.3--8 keV band
in units of $10^{-14}$ ergs cm$^{-2}$ s$^{-1}$.
}
\label{gamma-norm}
\end{figure}

To better constrain the model parameters, we have to fix the hydrogen
column density. Using the pulsar's dispersion measure 
(${\rm DM} = 2.38$ cm$^{-3}$ pc$^{-1}$, which corresponds to the electron
column density $N_e=7.3\times 10^{18}$ cm$^{-2}$) and assuming
the typical 10\% ISM ionization,
we obtain
$\nh= 7.3\times10^{19}$ cm$^{-2}$.
With the hydrogen column density fixed at this value, we obtain
a smaller photon index, $\Gamma\approx2.2$,
and a lower
unabsorbed flux, $F_{\rm 0.3-8\,keV}^{\rm unabs}=1.1\times 10^{-14}$
ergs cm$^{-2}$ s$^{-1}$ (see Fig.\ \ref{count_spectra}, {\em top}, 
Fig.\ \ref{gamma-norm}, and Table 1), corresponding to the isotropic luminosity
$L_{\rm 0.3-8\,keV}\approx 2.1\times 10^{28} d_{130}^2$ ergs s$^{-1}$
($d_{130}=d/130\,{\rm pc}$). Changing the value of 
the fixed hydrogen column density within a reasonable
range, $\nh \lesssim N_{\rm HI}$,
changes the fitting
parameters only slightly because the ISM absorption in the
observed energy band is very low at such small $\nh$ values
(see the fit for $\nh=2\times 10^{20}$ cm$^{-2}$ in Table 1).

\begin{figure}[h]
% \centering
\hspace{-0.3cm}
\includegraphics[width=2.6in,angle=90]{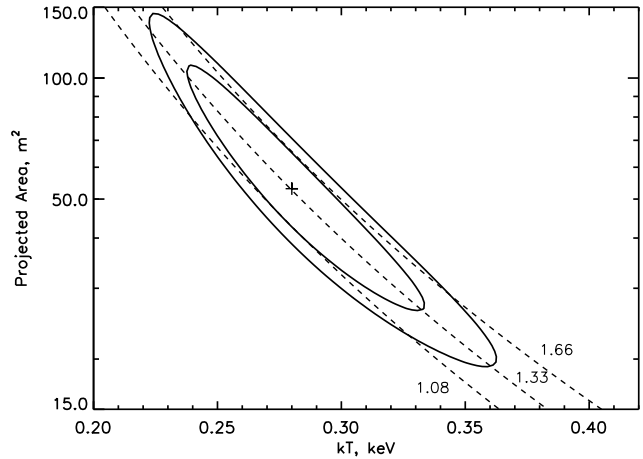}
\caption{Confidence contours (68\% and 90\%)
for the absorbed
BB model with
fixed $\nh = 7.3 \times10^{19}$ cm$^{-2}$.
The BB normalization is expressed in terms of projected
 emitting area,
in units of m$^{2}$ for $d=130$ pc.
The dashed lines are
the lines of constant bolometric luminosity,
in units of 10$^{28}$ ergs s$^{-1}$, assuming $d=130$ pc.
}
\vspace{0.4cm}
\label{kT-area}
\end{figure}

The fit of the same count-rate spectrum with the absorbed BB
 model with the hydrogen column density allowed to vary yields lower
$\nh$ values ($\nh < 5\times 10^{20}$ cm$^{-2}$, consistent with
$\nh= 7.3\times10^{19}$ cm$^{-2}$ estimated from the DM value;
see Fig.\ \ref{nh-gamma}, {\em bottom}, and Table 1).
The fit with
the column density fixed at $\nh= 7.3\times10^{19}$ cm$^{-2}$ 
gives approximately the same 
temperature
$kT\approx 
0.28$ keV ($T\approx 3.2$ MK) and
 projected emitting area $A_{\perp}
\sim 53 d_{130}^{2}$ m$^{2}$ as the fit with free $\nh$
(see Fig.\ \ref{count_spectra}, {\em bottom}, and Fig.\ \ref{kT-area}). 
This temperature and area correspond to the apparent radius
$R=(A_{\perp}/\pi)^{1/2}
\sim 4.1 d_{130}$ m
and bolometric luminosity $L_{\rm bol}=4A_{\perp}\sigma T^4
\sim 1.3\times 10^{28} d_{130}^2$ ergs s$^{-1}$.
Similar to the PL fit, the fitting parameters very weakly depend on
$\nh$ (see Table 1) as long as it remains within reasonable limits.
We also tried a PL+BB fit, but the fitting parameters could not be
constrained because of the small number of counts.

\section{Discussion}
\subsection{Proper motion}

%As we have mentioned in \S\,2.1, the position of the pulsar determined
%from the {\sl Chandra} ACIS image is different from those reported from
%the radio observations, which allows one to estimate the pulsar's
%proper motion. 
Based on the large X-ray/optical flux ratio, we have concluded in \S\,2.2 that
the point source we detected with the {\sl Chandra} ACIS is the X-ray
counterpart of the J0108 pulsar. As its X-ray position is
different from those reported from earlier radio observations (see \S\,2.2), 
we can estimate
the pulsar's proper motion.

Among the previously measured pulsar coordinates,
the most reliable ones, \\
$\alpha=01^{\rm h}08^{\rm m}08.317^{\rm s}\pm 0.010^{\rm s}$,
$\delta=-14^\circ 31' 49.35''\pm0.35''$,
 were obtained in the 11.3 hr interferometric observation
with the Australia Telescope Compact Array (ATCA) at epoch MJD 51,999
(Mignani et al.\  2003). Comparing these coordinates with those we
measured, we obtain the proper motion
\be
\mu_\alpha = 92\pm 44\,{\rm mas/yr}, \quad \mu_\delta=-176\pm 70\,{\rm mas/yr},
\ee
or
\be
\mu=199\pm 65\,{\rm mas/yr}, \quad {\rm P.A.}=
152^\circ\pm15^\circ
\ee
for the total proper motion and the position angle (counted east of north).

In addition to the interferometric position, timing positions of J0108
have been reported by Tauris et al.\ (1994), D'Amico et al.\ (1998), and
Hobbs et al.\ (2004). The position reported by Hobbs et al.
($\alpha=01^{\rm h}08^{\rm m}08.30^{\rm s}\pm 0.03^{\rm s}$,
$\delta=-14^\circ 31' 48.4\pm0.9''$; mean epoch MJD\,50,889 [1998.3]),
quoted in the current version of the ATNF Pulsar Catalogue
(Manchester et al.\ 2005), is supposed
to be the most accurate because it used data acquired during the longest 
time span of 9.6 yr, versus 1 yr and 3 yr for the measurements by
Tauris et al.\ and
D'Amico et al., respectively. Moreover, the Hobbs et al.\ position
  supresedes the earlier ones 
because it includes the data used by Tauris et al.\ and 
D'Amico et al.\footnote{ 
We should note that the position reported by D'Amico et al.\ is not
consistent with the other two timing positions if its very small 
errors ($0.005^{\rm s}$
and $0.1''$ in R.A.\ and decl., respectively) are
taken at face value. As these position errors are a factor of 10 smaller
than those of the Hobbs et al.\ position based on a factor of 3
longer observation time span, we suppose that the errors were strongly
underestimated. Therefore, the upper limit
 on the proper motion, $\mu < 82$ mas/yr,
obtained by Mignani et al.\ (2003) from the comparison of the ATCA position with the position
reported by D'Amico et al., should be disregarded.}
Using the Hobbs et al.\ position together with  the
the ATCA and {\sl Chandra}
positions,
we find the proper motion 
\be
\mu_\alpha = 91\pm 40\,{\rm mas/yr}, \quad \mu_\delta=-189\pm61\,{\rm mas/yr},
\ee
\be
\mu=210\pm57\,{\rm mas/yr}, \quad {\rm P.A.}=154^\circ\pm12^\circ ,
\ee
consistent with that given by equations (2) and (3). (We should note, however,
that the weights of the ATCA-{\sl Chandra} pair in this $\mu_\alpha$
and $\mu_\delta$ values, 0.76 and 0.73, respectively, are larger than
those of the two other pairs because of the
larger uncertainties of the timing position.)

The inferred proper motion corresponds to the transverse pulsar velocity
$V_\perp = (130\pm 35) d_{130}$ km s$^{-1}$, somewhat lower than the
typical transverse velocity of 300--400 km s$^{-1}$. 
The proper motion in the Galactic coordinates is 
$\mu_l\approx 140\,{\rm mas/yr}$, $\mu_b \approx -155\,{\rm mas/yr}$, i.e., the
pulsar is moving farther south from the Galactic plane.

Interestingly, a recent re-analysis of the original VLT observations
(epoch 2000.6) 
have shown a faint object 
($U=26.4\pm 0.3$, $B\approx 27.9$, $V\geq 27.8$), whose position 
(at the outskirts of the bright field galaxy) virtually
coincides with that following from
the backward extrapolation of the proper motion to the epoch
of the VLT observations (Mignani et al.\ 2008).
Assuming this is the optical counterpart of J0108,
its X-ray/optical flux ratios for the V and B bands
 ($R_V\geq 220$, $R_B\approx 160$) are consistent with typical values for 
radio pulsars (Zavlin \& Pavlov 2004; Zharikov et al.\ 2006), while the
relatively high flux in the U band might indicate thermal emission from
the bulk of the NS surface with a brightness temperature $\sim 9\times 10^4$ K,
too low to be detected in X-rays\footnote{Such a relatively high
surface temperature of the very old NS could be explained by (re)heating processes
in the NS interiors (see Mignani et al.\ 2008 for discussion and 
references).}. The optical detection (if
confirmed) provides additional support to the identification of the
detected X-ray source with the J0108 pulsar
 and to our estimate of the proper motion.

\begin{figure}[t]
% \centering
\hspace{-0.7cm}
\includegraphics[width=2.8in,angle=90]{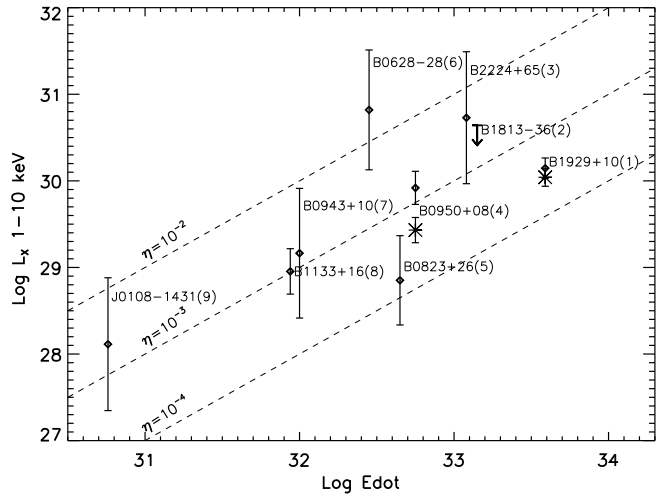}
\caption{
X-ray luminosities
of nine old pulsars versus spin-down
power. The numbers in parentheses correspond to the pulsars marked in Fig.\ 1.
For PSR B0950+08 and PSR 1929+10, the diamond and the asterisk show the
luminosities of the nonthermal and thermal components, respectively (as
determined by Zavlin \& Pavlov 2004 and Misanovic et al.\ 2008, respectively).
For the other pulsars, the luminosities were obtained from PL fits (see Fig.\ 5
in Kargaltsev et al.\ 2006). For the pulsars whose parallaxes have not
been measured,  we ascribe
a factor of 2  uncertainty to the distances estimated from the pulsar's
dispersion measure.
}
\label{Lx-Edot}
\end{figure}

\subsection{Energetics and spectrum}

J0108 is 
the least powerful and the oldest among the 
ordinary (nonrecycled) pulsars detected in the X-ray
range. The previous record-holders
have a factor of 15 higher spin-down power (B1133+16) and a factor of
9.5 smaller spin-down age (B0950+08).
Therefore, 
our observation of J0108 
has yielded the pulsar X-ray properties,
such as the spectrum and luminosity, in the yet
unexplored parameter domain, allowing us to examine the 
evolution of pulsar X-ray emission at very late stages of pulsar
activity.

The small number of source counts detected
 does not allow us to differentiate between
the alternative spectral models for J0108. 
For the PL fit with the hydrogen column density fixed at 
$\nh = 7.3\times 10^{19}$ cm$^{-2}$, we obtained 
the spectral slope $\Gamma=2.2\pm 0.2$, somewhat softer than 
the typical values, 
$\Gamma\sim1$--2, observed for young, more powerful pulsars (e.g.,
Gotthelf 2003). However, some old pulsars show even steeper spectra,
$\Gamma\sim 2$--3 (e.g. $\Gamma=2.5\pm 0.3$ for PSR B1133+16; 
Kargaltsev et al.\ 2006), 
suggesting that pulsar spectra might soften with
increasing age (i.e., decreasing spin-down power).
This softening can be ascribed to either an evolution of the 
 energy spectrum of magnetospheric particles or to a larger contribution
of soft thermal emissison from polar caps in old, low-powered
pulsars
(e.g., Zavlin \& Pavlov 2004). 

The luminosity found from the PL fit 
(e.g., 
$L_{\rm 0.3-10\, keV}\approx 2.1\times 10^{28}d_{130}^2$ ergs s$^{-1}$;
see \S\,2.2 and Table \ref{spectral})
is the lowest among the luminosities of ordinary pulsars detected in X-rays.
This is demonstrated in Figure \ref{Lx-Edot}, where we plot the
1--10 keV luminosities of nine old,
low-powered pulsars
($\edot < 10^{34}$ ergs s$^{-1}$).
Adding J0108 to the sample makes it clear that the X-ray luminosity
of old pulsars is correlated with the spin-down power in the same
manner as has been observed
for younger pulsars (i.e., $L_X$ generally grows with $\edot$,
albeit with a large scatter;
e.g., Possenti et al.\ 2002; Li et al.\ 2008;
 Kargaltsev \& Pavlov 2008).
The X-ray efficiency, $\eta_X=L_X/\edot$,
can be estimated for J0108 as
$\eta_{\rm 0.3-8\,keV}\sim 4\times 10^{-3}d_{130}^2$.
This efficiency is larger than the typical efficiencies, 
$\sim 10^{-5}$--$10^{-4}$, of younger, more powerful pulsars 
(see e.g., Fig.\ 5 in Kargaltsev \& Pavlov 2008),
but it is comparable with
the efficiencies of old pulsars (see Fig.\ \ref{Lx-Edot}). 
This supports the conjecture that,
on average,
old pulsars radiate a larger fraction of their spin-down power
in the X-ray range than younger ones (Zharikov et al.\ 2006;
Kargaltsev et al.\ 2006).

The higher X-ray efficiency could be attributed to the additional contribution
of soft X-ray polar cap emission, in accordance with the 
softer X-ray spectra of 
old pulsars. This hypothesis could be tested by deep
X-ray observations that would allow phase-resolved 
spectroscopy and energy-resolved timing (because the polar cap radiation,
which dominates at lower energies,
should show shallower pulsations than the magnetospheric radiation;
e.g., Zavlin \& Pavlov 2004).
Alternatively, one could speculate that the broadband 
magnetospheric spectrum shifts toward lower energies with decreasing
$\edot$, so that the soft X-ray efficiency increases at the expense
of decreasing hard X-ray or soft $\gamma$-ray efficiency.
Observations at higher energies with future missions could verify this 
hypothesis.

As we have mentioned in \S2.2,
the observed X-ray spectrum of J0108 can be fitted with a BB model,
which might be interpreted as thermal radiation from heated polar caps.
The best-fit BB temperature, 3.2 MK, obtained from this fit, is 
within the 1.7--3.5 MK range found from BB (or PL+BB) fits
for other old pulsars
(Zavlin \& Pavlov 2004; Kargaltsev et al.\ 2006; Misanovic et al.\ 2008;
Gil et al.\ 2008).
The projected emitting
 area, $A_\perp\sim 50\, d_{130}^2$ m$^{2}$, is, however,
exremely small.
Not only it is considerably smaller than the smallest area 
reported
($\sim 240$ m$^2$ for PSR B0834+06), but also the ratio, 
$\sim 6\times 10^{-4}d_{130}^2$,
of this area to the conventional polar cap area,
$A_{\rm pc}=2\pi^2 R_{\rm NS}^3/(cP)\sim 8.7\times 10^4$ m$^2$, 
is smaller than for any other
pulsar. We should bear in mind, however, that both the temperature
and the area are not only highly uncertain because of the strong correlation
of these parameters (see Fig.\ \ref{kT-area}), but 
they are also strongly 
dependent on the model of thermal emission (e.g., fits with hydrogen
NS atmosphere models usually give a factor of 2 lower 
effective temperature
and a factor of 10--100 larger area; Pavlov et al.\ 1995; 
Zavlin \& Pavlov 2004).
The large difference between the observed and conventional areas can
be also explained by the assumption that the 
non-dipolar component of the magnetic field at the NS
surface is much stronger than the dipolar component inferred from
the pulsar spindown, in accordance with the partially screened gap model for
the inner acceleration region above the pulsar polar cap (Gil et al.\ 2003).
Finally, the relationship between
 the actual emitting area and $A_\perp$ depends on
the orientation of the magnetic and spin axes, and on the General
Relativity effects (see discussion in Pavlov et al.\ 2007).

The apparent bolometric luminosity, 
$L_{\rm bol}\approx 1.3\times 10^{28} d_{130}^2$ ergs s$^{-1}$, which
is less model-dependent than $A_\perp$ and $T$ (Zavlin \& Pavlov 2004), 
is also smaller than those of the other old pulsars;
however, the corresponding 
``polar cap efficiency'', $\eta_{\rm pc}\equiv L_{\rm bol}/\edot
\sim 2\times 10^{-3}d_{130}^2$, is comparable to (or even larger than) those
reported for other pulsars. Such an efficiency 
exceeds the predictions
of the polar cap heating models by Harding \& Muslimov (2001, 2002)\footnote{Since
the curvature radiation cannot induce a pair cascade in the low-powered
J0108,
 only the less efficient inverse Compton
scattering (ICS)
 cascade can be responsible for polar cap heating in this pulsar.
The $L_{\rm bol}$ and $\eta_{\rm pc}$ estimated from our observation 
of J0108 are close to the analytical estimates for upper limits 
on these quantities for the resonant ICS heating (eqs.\ [60] and [64] in
Harding \& Mulsimov 2002). However, 
an extrapolation of the numerical calculations
of these authors into the J0108 parameter domain seems to predict lower
values.
}, 
but it is in a reasonable agreement with the partially screened gap 
model by Gil et al.\ (2003, 2008) .
We should not forget, however, that this $\eta_{\rm pc}$
 is only an upper limit because
%it is hardly plausible that the observed emission is purely thermal.
%(At least, the spectra of the brighter pulsars B1929+10 and B0950+08 certainly
%contain nonthermal components.)
the observed spectrum can contain a nonthermal component.
% Thus, we cannot rule out the presence
%of a thermal component in the spectrum of J0108, but the data available
%do not allow us to separate it from the magnetospheric emission.

\subsection{Summary}

Thanks to the high sensitivity of the {\sl Chandra} ACIS, we
have detected J0108,
 the oldest and the least powerful pulsar ever observed in X-rays.
This observation has shown that even very old pulsars can be
rather efficient X-ray emitters, even more efficient than young ones. 
It seems likely that the observed emission
contains both magnetospheric and thermal components. To separate these
components and understand the nature of the X-ray emission, 
a deep \xmm\ observation, which would provide not only
high sensitivity but also a sufficient time resolution, would be particularly
useful. 
To evaluate the X-ray luminosity and efficiency of J0108 more
accurately, the distance to the pulsar should be determined from
annual parallax measurements.

Comparing the {\sl Chandra} position of J0108 with the previously measured
radio positions, we have been able to estimate the pulsar's proper motion of
about 200 mas/yr. To determine the proper motion with a higher accuracy,
radio observations with ATCA and/or VLA are required. Alternatively,
the proper motion can be accurately measured with new \chan\
observations in a few years from now.

The detection of the candidate optical counterpart of J0108 (Mignani
et al.\ 2008) was hampered by the proximity of its sky position
 to the bright field galaxy
%The detection of J0108 in the X-ray band strongly increases the chance
%to detect it in the optical\footnote{
%It is plausible that the optical counterpart of the pulsar 
%could have been detected in the VLT observation by Mignani et al. (2003)
%if it were not projected so close to the nearby field galaxy 
(see Fig.\ 3).
% \ref{VLT_image}). 
Our estimate of the proper motion suggests that
it has moved away from the galaxy and could be firmly detected in
the near future.
%Based on the correlation between the X-ray and optical fluxes
%of magnetospheric radiation (Zavlin \& Pavlov 2004; Zharikov et
%al.\ 2004), we can expect optical magnitudes of about 28--30 for the pulsar 
%counterpart. 
Measuring its optical-UV spectrum would be particularly important for
understanding the thermal evolution and magnetospheric emission of old NSs.

\acknowledgements
We thank Roberto Mignani for providing the reduced VLT images
of the pulsar field and useful discussions. 
This work was partially supported by 
\chan\ award SV4-74018.

\end{document}